\newcommand{\ie}{{\it i.e.}}
\newcommand{\eg}{{\it e.g.}}

\newcommand{\etal}{{\it et al.}}
\newcommand{\lhs}{l.h.s.}
\newcommand{\rhs}{r.h.s.}
\newcommand{\BR}[1]{\linebreak[0]#1\linebreak[0]}

\documentstyle[prb,aps]{revtex}

\draft

\begin{document}

\title{
  Bunching of fluxons by the Cherenkov radiation in
  Josephson multilayers
}

\author{
  E.~Goldobin\cite{gold-mail}
}

\address{
  Institute of Thin Film and Ion Technology (ISI),
  Research Center J\"ulich GmbH (FZJ) \\
  D-52425 J\"ulich, Germany
}

\author{
  B.~A.~Malomed\cite{Boris-mail}
}

\address{
  Department of Interdisciplinary Studies, Faculty of
  Engineering\\
  Tel Aviv University, Tel Aviv 69978, Israel
}

\author{
  A.~V.~Ustinov
}

\address{
  Physikalisches Institut III,
  Uni\-ver\-si\-t\"at Er\-lan\-gen-N\"urn\-berg,
  D-91054, Erlangen, Germany
}

\date{\today}

\wideabs{ 

\maketitle

\begin{abstract}

   A single magnetic fluxon moving at a high velocity in a Josephson
   multilayer (\eg, high-temperature superconductor such as BSCCO) can
   emit electromagnetic waves (Cherenkov radiation), which leads to
   formation of novel stable dynamic states consisting of several bunched
   fluxons. We find such bunched states in numerical simulation in the
   simplest cases of two and three coupled junctions. At a given driving
   current, several different bunched states are stable and move at
   velocities that are higher than corresponding single-fluxon velocity.
   These and some of the more complex higher-order bunched states and
   transitions between them are investigated in detail.

\end{abstract}


} 

\section{Introduction}

In the recent years, a great deal of attention was attracted
to different kinds of solid state multilayered systems, \eg,
artificial Josephson and magnetic multilayers, high-temperature 
superconductors (HTS) and perovskites, to name
just a few. Multilayers are attractive because it is often
possible to multiply a physical effect achieved in one layer
by $N$ (and sometimes by $N^2$), where $N$ is the number of
layers. This can be exploited for fabrication of novel 
solid-state devices. In addition, multilayered solid state systems
show a variety of new physical phenomena which result from
the interaction between individual layers.

In this article we focus on Josephson multilayer, the
simplest example of which is a stack consisting of just two
long Josephson junctions (LJJs). The results of our
consideration can be applied to intrinsically layered HTS
materials \cite{Intrinsic}, since the
Josephson-stack model has proved
to be appropriate for these structures
\cite{Kl-Mue-Kolh:94,Paul:AdvSSP34,pl:pap154}.

In earlier papers \cite{image,Cherry1,Cherry2,HexFisherCherry} it
was shown that, in some cases, a fluxon (Josephson vortex)
moving in one of the layers of the stack may emit
electromagnetic (plasma) waves by means of the Cherenkov
mechanism. The fluxon together with its Cherenkov radiation
have a profile of a traveling wave, $\phi(x-ut)$, having an
oscillating gradually decaying tail. Such a wave profile
generates an effective potential for another fluxon which
can be added into the system. If the second fluxon is
trapped in one of the minima of this traveling potential, we
can get a {\it bunched state} of two fluxons. In such a state, two
fluxons can stably move at a small constant distance one from
another, which is not possible otherwise.
Fluxons of the same polarity usually repel each other, even
being located in different layers.

Similar bunched states were already found in a discrete
Josephson transmission lines\cite{Ustinov:BunchArray}, as
well as in long Josephson junctions with the so-called
$\beta$-term due to the surface impedance of the
superconductor
\cite{Sakai:BunchBeta,Malomed:BunchBeta,Vernik:BunchBeta}.
The dynamics of conventional LJJ is described by the sine-
Gordon equation which does not allow the fluxon to move
faster than the Swihart velocity and, therefore, the Cherenkov
radiation never appears. In both cases mentioned above (the
discrete system or the system with the $\beta$-term), the
perturbation of the sine-Gordon equation results in a modified
dispersion relation for Josephson plasma waves and
appearance of an oscillating tail. This tail, in turn,
results in an attractive interaction between fluxons, \ie,
bunching. Nevertheless, the mere presence of an oscillating
tail is not a sufficient condition for bunching.


In this paper, we investigate the problem of fluxon bunching
in a system of two and three inductively coupled junctions
with a primary state $[1|0]$ (one fluxon in the top junction and
no fluxon in the bottom one) or $[0|1|0]$ (a fluxon only in the
middle junction of a 3-fold stack). We show that bunching is
possible for some fluxon configurations and specific range
of parameters of the system. In addition, it is found that
the bunched states radiate less than single-fluxon states,
and therefore can move with a higher velocity. Section
\ref{Sec:SimRes} presents the results of numerical
simulations, in section \ref{Sec:Discussion} we discuss the
obtained results and a feasibility of experimental
observation of bunched states. We also derive a simple
analytical expression which show the possibility of the
existence of bunched states. Section \ref{Sec:Conclusion}
concludes the work.

\section{Numerical Simulations}
\label{Sec:SimRes}

The system of equations which describes the dynamics of
Josephson phases $\phi^{A,B}$ in two coupled LJJ$^A$ and
LJJ$^B$ is \cite{SBP,LT21}:
\begin{eqnarray}
  \frac{\phi^A_{xx}}{1-S^2}
  - \phi^A_{tt}
  - \sin{\phi^A}
  - \frac{S}{1-S^2} \phi^B_{xx}
  &=&\alpha \phi^A_{t} - \gamma
  \ ; \label{Eq:2:PDEa}\\
  \frac{\phi^B_{xx}}{1-S^2}
  -  \phi^B_{tt}
  -  \frac{\sin{\phi^B}}{J}
  -   \frac{S}{1-S^2} \phi^A_{xx}
  &=& \alpha \phi^B_{t} - \gamma
  \ , \label{Eq:2:PDEb}
\end{eqnarray}
where $S$ ($-1<S<0$) is a dimensionless coupling constant, $J=j_c^A/j_c^B$ is the ratio of the critical currents, while $\alpha$ and $\gamma=j/j_c^A$ are the damping coefficient and normalized bias current, respectively, that are assumed to be the same in both LJJs. It is also assumed that other parameters of the junctions, such as the effective magnetic thicknesses and capacitances, are the same. As has been shown earlier\cite{Cherry1,Cherry2}, the Cherenkov radiation in a two-fold stack may take place only if the fluxon is moving in the junction with smaller $j_c$. We suppose in the following that the fluxon moves in LJJ$^A$, that implies $J<1$.

In the case $N=3$, we impose the symmetry condition $\phi^{A}\equiv \phi^{C}$, which is natural when the fluxon moves in the middle layer, and, thus, we can rewrite equations from Ref.~[\onlinecite{SBP}] in the form
\begin{eqnarray}
  \frac{\phi^A_{xx}}{1-2S^2} - \phi^A_{tt} -\sin\phi^A
  -\frac{S\phi^B_{xx}}{1-2S^2} &=&\alpha \phi^A_t - \gamma
  \ ; \label{Eq:Sym3:AC} \\
  \frac{\phi^B_{xx}}{1-2S^2}-\phi^B_{tt}-\sin\phi^B
  -\frac{2S\phi^A_{xx}}{1-2S^2} &=&\alpha \phi^B_t - \gamma
  \ .  \label{Eq:Sym3:B}
\end{eqnarray}
Note the factor 2 in the last term on the \lhs{} of Eq.~(\ref{Eq:Sym3:B}). In the case of three coupled LJJs, we assume $J=1$, since for more than two coupled junctions the Cherenkov radiation can be obtained for a uniform stack with equal critical currents\cite{Cherry2}

\subsection{Numerical technique}

The numerical procedure works as follows. For a given set of the LJJs parameters, we compute the current-voltage characteristic (IVC) of the system, \ie, $\bar{V}^{A,B}(\gamma)$. To calculate the voltages $\bar{V}^{A,B}$ for fixed values of $\gamma$, we simulate the dynamics of the phases $\phi^{A,B}(x,t)$ by solving Eqs.~(\ref{Eq:2:PDEa}) and (\ref{Eq:2:PDEb}) for $N=2$ or Eqs.~(\ref{Eq:Sym3:AC}) and (\ref{Eq:Sym3:B}) for $N=3$, using the periodic boundary conditions:
\begin{eqnarray}
  \phi^{A,B}(x=L) &=& \phi^{A,B}(x=0) + 2\pi N^{A,B}
  ; \label{Eq:BC:phi}\\
  \phi_x^{A,B}(x=L) &=& \phi_x^{A,B}(x=0)
  , \label{Eq:BC:phi'}
\end{eqnarray}
where $N^{A,B}$ is the number of fluxons trapped in LJJ$^{A,B}$. In order to simulate a quasi-infinite system, we have chosen annular geometry with the length (circumference) of the junction $L=100$.

To solve the differential equations, we use an explicit method [expressing $\phi^{A,B}(t+\Delta t)$ as a function of $\phi^{A,B}(t)$ and $\phi^{A,B}(t-\Delta t)$], treating $\phi_{xx}$ with a five-point, $\phi_{tt}$ with a three-point, and $\phi_{t}$ with a two-point symmetric finite-difference scheme. The spatial and time steps used for the simulations were $\delta x = 0.025$, $\delta t=0.00625$. After the simulation of the phase dynamics for $T=10$ time units, we calculate the average dc voltages $\bar{V}^{A,B}$ for this time interval as
\begin{equation}
  \bar{V}^{A,B}
  = \frac{1}{T}\int_0^T \phi^{A,B}_t(t) \:dt
  = \frac{\phi^{A,B}(T)-\phi^{A,B}(0)}{T}
  \quad . \label{Eq:V}
\end{equation}
The dc voltage at point $x$ can be defined as average number of fluxons (the flux) passed through the junction at this point. Since the average fluxon density is not singular in any point of the junction (otherwise the energy will grow infinitely), we conclude that average dc voltage is the same for any point $x$. Therefore, for faster convergence of our averaging procedure, we can additionally average the phases $\phi^{A,B}$ in (\ref{Eq:V}) over the length of the stack.

After the values of $\bar{V}^{A,B}$ were found as per Eq. (\ref{Eq:V}), the evolution of the phases $\phi^{A,B}(x,t)$ is simulated further during $1.1\:T$ time units, the dc voltages $\bar{V}^{A,B}$ are calculated for this new time interval and compared with the previously calculated values. We repeat such iterations further, increasing the time interval by a factor 1.1 until the difference in dc voltages $|\bar{V}(1.1^{n+1}\:T)-\bar{V}(1.1^n\:T)|$ obtained in two subsequent iterations becomes less than an accuracy $\delta V=10^{-4}$. The particular factor $1.1$ was found to be quite optimal and to provide for fast convergence, as well as more efficient averaging of low harmonics on each subsequent step. Very small value of this factor, \eg, $1.01$ (recall that only the values greater than 1 have meaning), may result in a very slow convergence in the case when $\phi(t)$ contains harmonics with the period $\ge{}T$. Large values of the factor, \eg, $\ge{}2$, would consume a lot of CPU time already during the second or third iteration and, hence, are not good for practical use.

Once the voltage averaging for current $\gamma$ is complete, the current $\gamma$ is increased by a small amount $\delta\gamma = 0.005$ to calculate the voltages at the next point of the IVC. We use a distribution of the phases (and their derivatives) achieved in the previous point of the IVC as the initial distribution for the following point.

Further description of the software used for simulation can be found in Ref.~\onlinecite{StkJJ}.

\subsection{Two coupled junctions}

For simulation we chose the following parameters of the system: $S=-0.5$ to be close to the limit of intrinsically layered HTS, $J=0.5$ to let the fluxon accelerate above the ${\bar c}_{-}$ and develop Cherenkov radiation tail. The velocity ${\bar c}_{-}$ is the smallest of Swihart velocities of the system. It characterizes the propagation of the out-of-phase mode of Josephson plasma waves. The value of $\alpha=0.04$ is chosen somewhat higher than, \eg, in (Nb-Al-AlO$_x$)$_N$-Nb stacks. This choice is dictated by the need to keep the quasi-infinite approximation valid and satisfy the condition $\alpha L \gg 1$. Smaller $\alpha$ requires very large $L$ and, therefore, unaffordably long simulation times. So, we made a compromise and chose the above $\alpha$ value.

First, we simulated the IVC $u(\gamma)$ in the $[1|0]$ state, by sweeping $\gamma$ from 0 up to 1 and making snapshots of phase gradients at every point of the IVC. This IVC is shown in Fig.~\ref{Fig:IVC+Profile[1|0]}(a), and the snapshot of the phase gradient at $\gamma=0.3$ is presented in Fig.~\ref{Fig:IVC+Profile[1|0]}(b). As one can see, the Cherenkov radiation tail, which is present for $u>{\bar c} _{-}$, has a sequence of minima where the second fluxon may be trapped.

\subsubsection{$[1+1|0]$ state}

In order to create a two-fluxon bunched state and check its stability, we used the following ``solution-engineering'' procedure. By taking a snapshot of the phase profiles  $\phi_{A,B}(x)$ at the bias value $\gamma_0=0.3$, we constructed an {\it ansatz} for the bunched solution in the form
\begin{equation}
  \phi_{A,B}^{\rm new}(x) = \phi_{A,B}(x) + \phi_{A,B}
  (x+\Delta{}x)
  , \label{Eq:NewTrialFn[2|0]}
\end{equation}
where $\Delta{}x$ is chosen so that the center of the trailing fluxon is placed at one of the minima of the Cherenkov tail. For example, to trap the trailing fluxon in the first, second and third well, we used $\Delta{}x=0.9$, $\Delta{}x= 2.4$ and $\Delta{}x=3.9$, respectively. The phase distribution (and derivatives), constructed in this way, were used as the initial condition for solving Eqs.~(\ref{Eq:2:PDEa}) and (\ref{Eq:2:PDEb}) numerically. As the system relaxed to the desired state $[1+1|0]$, we further traced $u(\gamma)$ curve, varying $\gamma_0$ down to $0$ and up to $1$.

We accomplished this procedure for a set of $\Delta{}x$ values, trying to trap the second fluxon in every well. Fig.~\ref{Fig:IVC+Profile[1|0]}(c) shows that a stable, tightly bunched state of two fluxons is indeed possible. Actually, all the $[1+1|0]$ states obtained this way have been found to be stable, and we were able to trace their IVCs up and down, starting from the initial value of the bias current $\gamma=0.3$. For the case when the trailing fluxon is trapped in the first, second and third minima, such IVCs are shown in Fig.~\ref{Fig:IVC[2|0]}.

The most interesting feature of these curves is that they correspond to the velocity of the bunched state that is {\em higher}\/ than that of the $[1|0]$ state, at the same value of the bias current. Comparing solutions shown in Figs.~\ref{Fig:IVC+Profile[1|0]}(b) and \ref{Fig:IVC+Profile[1|0]}(c), we see that the amplitude of the trailing tail is smaller for the bunched state. This circumstance suggests the following explanation to the fact that the observed velocity is higher in the state $[1+1|0]$ than in the single-fluxon one. Because the driving forces acting on two fluxons in the bunched and unbunched states are the same, the difference in their velocities can be attributed only to the difference in the friction forces. The friction force acting on the fluxon in one junction is
\begin{equation}
  F_{\alpha} = \alpha \int_{-\infty}^{+\infty}
  \phi_x\phi_t\:dx
  \, , \label{Eq:Falpha}
\end{equation}
and the same holds for the other junction. By just looking at Fig.~\ref{Fig:IVC+Profile[1|0]}(b) and (c) it is rather difficult to tell in which case the friction force is larger, but accurate calculations using Eq.~(\ref{Eq:Falpha}) and profiles from Fig.~\ref{Fig:IVC+Profile[1|0]}(b) and (c) show that the friction force acting on two fluxons with the tails shown in Fig.~\ref{Fig:IVC+Profile[1|0]}(b) is somewhat higher than that for Fig.~\ref{Fig:IVC+Profile[1|0]}(c). This result is not surprising if one recalls that, to create the bunched state, we have shifted the $[1|0]$ state by about half of the tail oscillation period relative to the other single-fluxon state. Due to this, the tails of the two fluxons add up out of phase and partly cancel each other, making the tail's amplitude behind the fluxon in the bunched state lower than that in the $[1|0]$ state.

From Fig.~\ref{Fig:IVC[2|0]} it is seen that every bunched state exists in a certain range of values of the bias current. If the current is decreased below some threshold value, fluxons dissociate and start moving apart, so that the interaction between them becomes exponentially small. When the trailing fluxon sits in a minimum of the Cherenkov tail sufficiently far from the leading fluxon, the IVC corresponding to this bunched state is almost undistinguishable from that of the $[1|0]$ state, as the two fluxons approach the limit when they do not interact. We have found that IVCs for $M>3$, where $M$ is the potential well's number, is indeed almost identical to that of the $[1|0]$ state. In contrast to bunching of fluxons in discrete LJJ\cite{Ustinov:BunchArray}, the transitions from one bunched state to another with different $M$ do {\emph not} take place in our system. Thus, we can say that the current range of a bunched state with smaller $M$ ``eclipses'' the bunched states with larger $M$.

The profiles of solutions found for various values of the bias current are shown in Fig.~\ref{Fig:Profiles[2|0]}. We notice that at the bottom of the step corresponding to the bunched state the radiation tail is much weaker and fluxons are bunched tighter. This is a direct consequence of the fact that at lower velocities the radiation wavelength and the distance between minima becomes smaller, and so does the distance between the two fluxons. At a low bias current, the radiation wavelength and, hence, width of the potential wells become very small and incommensurable with the fluxon's width. Therefore, the fluxon does not fit into the well and the bunched states virtually disappear.

\subsubsection{$[1|1]$ state}

The initial condition for this state was constructed in a
similar fashion to the $[1+1|0]$ one, but now using a
cross-sum of the shifted and unshifted solutions:
\begin{equation}
  \phi_{A,B}^{\rm new}(x) = \phi_{A,B}(x) + \phi_{B,A}
  (x+\Delta{}x)
  . \label{Eq:NewTrialFn[1|0+1]}
\end{equation}
If for the $[1+1|0]$ state, $\Delta{}x$ was $\approx(\lambda-\frac{1}{2})M$, $M=1,2\ldots$, then in the $[1|1]$ state we have to take $\Delta{}x\approx\lambda{}M$. We can also take $M=0$ \, i.e., $\Delta{}x=0$, which corresponds to the degenerate case of the in-phase $[1|1]$ state. The stability of this state was investigated in detail analytically by Gr{\o}nbech-Jensen and co-authors\cite{GrE:Stability}, and is outside the scope of this paper.

Our efforts to create a bound state $[1|1]$ using the phase
in the form (\ref{Eq:NewTrialFn[1|0+1]}) with $M=1,2\ldots$
have {\it not} lead to any stable configuration of bunched fluxons
with $\Delta x \ne 0$. 


\subsubsection{Higher-order states}

Looking at the phase gradient profiles shown in
Fig.~\ref{Fig:Profiles[2|0]}, one notes that these profiles
are qualitatively very similar to the original profile of
the soliton with a radiation tail behind it [see
Fig.~\ref{Fig:IVC+Profile[1|0]}(b)], with the only difference that
there are two bunched solitons with a tail. So, we can try
to construct two pairs of bunched fluxons moving together,
\ie, get a $[2+2|0]$ bunched state. As before, the trapping
of the trailing pair is possible in one of the minima of the
tail generated by the leading pair. To construct such a
double-bunched state we employ the initial conditions obtained
using Eq.~(\ref{Eq:NewTrialFn[2|0]}) at the bias point
$\gamma_0=0.3$, using the steady phase distribution obtained
for the $[2|0]$ state at $\gamma_0=0.3$. The shift $\Delta{}
x$ was chosen in such a way that a pair of fluxons fits into
one of the minima of the tail. We note that in this case we
needed to vary $\Delta{}x$ a little bit before we have
achieved trapping of the trailing pair in a desired well.

Simulations show that the obtained $[2+2|0]$ states are
stable and demonstrate an even {\it higher} velocity of the whole
four-fluxon aggregate. The corresponding IVCs and
profiles are shown in Fig.~\ref{Fig:IVC[2+2|0]}(a) and (b),
respectively. Note that at $\gamma<0.22$ the bunched state
$[2+2|0]$ splits first into $[1+1_2+1_3+1_3|0]$ state (the
subscripts denote the well's number $M$, counting from the
previous fluxon), and at still lower bias current,
$\gamma<0.2$, they split into two
independent $[1+1_2|0]$ and $[1+1_5|0]$ states. This two
states move with slightly different velocities and can
collide with each other due to the periodic nature of the
system. As a result of collisions, these states ultimately
undergo a transformation into two independent $[1+1_5|0]$
states. As the bias decreases below $\approx 0.1$, the
velocity $u$ becomes smaller than ${\bar c}_{-}$ and the
Cherenkov radiation tails disappear. At this point, each of
the $[1+1_5|0]$ states smoothly transforms into two
independent $[1|0]$ states. The interaction between these
states is exponentially small, with a characteristic length
$\sim 1$ (or, $\lambda_J$ in physical units). We note that
the interaction between kinks in the region $u>{\bar c}_{-}
$, where they have tails, also decreases exponentially, but
with a larger characteristic length $\sim\alpha^{-1}$.

The procedure of constructing higher-order bunched states
can be performed using {\em different} states as ``building
blocks''. In particular, we also tried to form the $[2+1|0]$ bunched
state. Note that if two different states are taken as
building blocks, we need to match their velocities, and,
hence, the wave lengths of the tail. Thus, we have
to combine two states at the same velocity, rather than at
the same bias current. Since different states have their own
velocity ranges, it is not always possible. As an example,
we have constructed a $[2+1|0]$ state out of $[2|0]$ state
at $\gamma=0.15$ and $[1|0]$ state at $\gamma=0.45$ using
an {\it ansatz} similar to (\ref{Eq:NewTrialFn[2|0]}). These states
have approximately the same velocity $u\approx0.95$ (see
to Fig.~\ref{Fig:IVC[2|0]}). The constructed state was
simulated, starting from the points $\gamma=0.3$ and $\gamma=
0.35$, tracing IVC up and down as before. Depending on the
bias current the system ends up in different states, namely
in the state $[1+1_1+1_2|0]$ for $\gamma_0=0.3$, or in the
state $[1+1_1+1_1|0]=[3|0]$ for $\gamma_0=0.35$. The IVCs of
both states are shown in Fig.~\ref{Fig:IVC[2+2|0]}. The
profiles of the phase gradients are shown in
Fig.~\ref{Fig:IVC[2+2|0]}(c).

Our attempts to construct the states with a higher number of bunched fluxons, e.g., $[4+4|0]$, have failed since four fluxons do not fit into one well. We have concluded that such states immediately get converted into one of the lower-order states.

\subsection{Three coupled junctions}

We have performed numerical simulation of
Eqs.~(\ref{Eq:Sym3:AC}) and (\ref{Eq:Sym3:B}), using the
same technique as described in the previous section. Our
intention here is to study the 3-junction case in which the
fluxon is put in the middle junction ($[0|1|0]$ state). All
other parameters were the same as in the case of the
two-junction system, except for the ratio of the critical
currents $J$, which was taken equal to one. This simplest
choice is made because in a system of $N>2$ coupled {\em
identical} junctions the Cherenkov radiation appears in a
$[0|\ldots|0|1|0|\ldots|0]$ state for $u>{\bar c}_{-}
\approx0.765$ (this pertains to $S=-0.5$).

Fig.~\ref{Fig:IVC[0|2|0]} shows the IVCs of the original
state $[0|1|0]$, as well as IVCs of the bunched state
$[0|1+1|0]$ for $M=1,\:2,\:3$. The profiles of the phase
gradients at points A through D are shown in
Fig.~\ref{Fig:Profiles[0|2|0]}. Qualitatively, the bunching
in the 3-fold system takes place in a similar fashion as
that in the 2-fold system. Nevertheless, we did not succeed
in creating a stable fluxon configuration with $M=3$,
although the stable states with other $M$ were obtained. We
would like to mention, that when the second fluxon was put
in the second minimum of the potential to get the state with
$M=2$, the state with $M=1$ has been finally
established as a result of relaxation. The same behavior was observed
when we put the fluxon initially in the third minimum, the
system ended up in the state $[1+1_2|0]$. For $M\ge4$, the
behavior was as usual. We tried to vary $\Delta{}x$ smoothly, so
that the center of the trailing fluxon would correspond to
different positions between the second and fourth well, but
in this case we did not succeed to get $[1+1_2|0]$ state.

Following the same way as for two coupled junctions, we
tried to construct $[0+1|1|0+1]$ states. As in the case $N=
2$, these states were found unstable for any $M>0$, e.g.,
they would split
into $[0|1+1_2|0]$ and $[1|-1|1]$.
The state $[0|2+2|0]$ was not stable either for $M=1,\:2,\:3$ and the
bias currents $\gamma_0=0.20$, 0.30, 0.35.

The state $[0|2+1|0]=[0|3|0]$, constructed by combining the solutions for the $[0|1|0]$ and $[0|2|0]$ states moving with equal velocities was found to be stable when starting at $\gamma=0.25$ and sweeping bias current up and down. The dependence $u(\gamma)$ is shown in Fig.~\ref{Fig:IVC[0|2|0]}. One may note, that for the states $[0|2|0]$ and $[0|3|0]$ the dependence is not smooth. Indeed, for these states the Cherenkov radiation tail is so long ($\sim L$), that our annular system cannot simulate an infinitely long system, resulting in Cherenkov resonances which inevitably appear in the system with a finite perimeter\cite{Cherry1,Cherry2}.

\section{Analysis and Discussion}
\label{Sec:Discussion}

Because of the non-linear nature of the bunching problem, it is
hardly tractable analytically. Therefore, we here present an
approach in which we analyze the asymptotic behavior of the
fluxon's front and trailing tails in the linear
approximation. This technique is similar to that employed
in Ref.~\onlinecite{Ustinov:BunchArray}. We assume that, at
distances which are large enough in comparison with the
fluxon's size, the fluxon's profile is exponentially decaying,
\begin{equation}
  \phi(x,t) \propto \exp[p(x-ut)]
  \quad , \label{Eq:tail}
\end{equation}
where $p$ is a complex number which can be found by
substituting this expression into Eqs.~(\ref{Eq:2:PDEa}) and
(\ref{Eq:2:PDEb}). As a result we arrive at an equation
\begin{equation}
  \left|
    \begin{array}{cc}
    \frac{p^2}{1-S^2}-p^2u^2-1-\alpha{}pu & -\frac{Sp^2}{1-
    S^2}\\
    -\frac{Sp^2}{1-S^2} & \frac{p^2}{1-S^2}-p^2u^2-\frac{1}
    {J}-\alpha{}pu
    \end{array}
  \right|=0
  \quad , \label{Eq:f(p)=0}
\end{equation}
In general, this yields a 4-th order algebraic equation which
always has 4 roots. If we want to describe a soliton moving
from left to right with a radiation tail behind it, we have
to find the values $p$ among the four roots which adequately describe
the front and rear parts of the soliton. Because the front (right)
part of the soliton is not oscillating, it is described by
Eq.~(\ref{Eq:tail}) with real $p<0$. The rear (left) part of the
soliton is the oscillating tail, consequently it should be described
by Eq.~(\ref{Eq:tail}) with complex $p$ having ${\rm Re}(p)>0$,
the period of oscillations being determined by the imaginary part of
$p$. Analyzing the 4-th order equation, we conclude that the two 
necessary types
of the roots coexist only for $u>{\bar c}_{-}$,
which is quite an obvious result.

To analyze the possibility of bunched state formation, we
consider two fluxons situated at some distance from each
other. We propose the following two conditions for the two
fluxons to form a bunched state:
\begin{enumerate}

  \item Since non-oscillating tails result only in repulsion
  between fluxons, while the oscillating tail leads to
  mutual trapping, the condition
  \begin{equation}
    {\rm{}Re}(p_l)<|p_r|
    \quad , \label{Eq:BunchCond1}
  \end{equation}
  can be imposed to secure bunching. Here $p_l$ is the root of
  Eq.~(\ref{Eq:f(p)=0}) which describes the left (oscillating)
  tail of the leading (right) fluxon, and $p_r$ is the root
  of Eq.~(\ref{Eq:f(p)=0}) which describes the right (non-oscillating) 
  tail of the trailing (left) fluxon.

  \item The relativistically contracted fluxon must fit into
  the minimum of the tail, i.e.,
  \begin{equation}
    \frac{\pi}{{\rm Im}(p)}>\sqrt{\frac{u^2}{{\bar c}_{-}
    ^2}-1}
    \quad , \label{Eq:BunchCond2}
  \end{equation}
  where $\pi/{\rm Im}(p)$ is half of the wavelength of the
  tail-forming
  radiation (the well's width), and the expression on
  the \rhs{} of Eq. (\ref{Eq:BunchCond2}) approximately
  corresponds to the contraction of the fluxon at the
  trans-Swihart velocities. Although our system is not Lorentz
  invariant, numerical simulations show that the fluxon
  indeed shrinks (not up to zero) when approaching the
  Swihart velocity ${\bar c}_{-}$ from both sides.

\end{enumerate}
Following this approach, we have found that the second condition
(\ref{Eq:BunchCond2}) is always satisfied. The first
condition (\ref{Eq:BunchCond1}) gives the following result.
Bunching is possible at $u>u_b>{\bar c}_{-}$. The value of
$u_b$ can be calculated numerically and for $S=-0.5$, $J=
0.5$, $\alpha=0.04$ it is $u_b=0.837$. Looking at
Fig.~\ref{Fig:IVC[2|0]}, we see that this velocity
corresponds to the bias point where the $[1+1_M|0]$ states
cease to exist. Thus, our crude approximation
reasonably predicts the velocity range where the
bunching is possible.

\section{Conclusion}
\label{Sec:Conclusion}

In this work we have shown by means of numerical simulations
that:
\begin{itemize}

  \item The emission of the Cherenkov plasma waves by a fluxon moving with high velocity creates an effective potential with many wells, where other fluxons can be trapped. This mechanism leads to bunching between fluxons of the {\em same} polarity.

  \item We have proved numerically that in the system of two and three coupled junctions the bunched states for the fluxons in the {\em same} junction such as $[1+1|0]$, $[1+2|0]$, $[2+2|0]$, $[0|1+1|0]$ are stable. The states with fluxons in different junctions like $[1|0+1]$ and $[0+1|1|0+1]$ are numerically found unstable (except for the degenerated case $M=0$, when $[1|1]$ is a simple in- phase state).

  \item Bunched fluxons propagate at a substantially higher velocity than the corresponding free ones at the same bias current, because of lower losses per fluxon.

  \item When decreasing the bias current, transitions between the bunched states with different separations between fluxons were not found. This behavior differs from what is known for the bunched states in a discrete system\cite{Ustinov:BunchArray}. In addition, a splitting of multi-fluxon states into the states with smaller numbers of bunched fluxons is observed.

\end{itemize}

\acknowledgments

This work was supported by a grant no. G0464-247.07/95 from the German-Israeli Foundation.


\begin{figure}
  \caption{
    (a) The current-velocity characteristic $u(\gamma)$ for
    the fluxon moving in the $[1|0]$ state (from left to
    right). (b) The profiles of the phase gradients
    $\phi^{A,B}_x(x)$ in the state $[1|0]$ at $\gamma=0.3$,
    corresponding to the bias point A shown in Fig.~(a). The
    Cherenkov tail, present at $u>{\bar c}_{-} \approx
    0.817$, has a set of minima where the second fluxon can
    be trapped. (c) The profiles of $\phi^{A,B}_x(x)$ in the
    state $[2|0]$ at the same value $\gamma=0.3$ as (b). Two
    fluxons shown in Fig.~(c) are almost undistinguishable.}
  \label{Fig:IVC+Profile[1|0]}
\end{figure}
\begin{figure}
  \caption{
    Current-velocity characteristics of different bunched
    states $[2|0]$: the second fluxon is trapped in the
    first minimum of the tail (state $[1+1_1|0]$), the second
    minimum (state $[1+1_2|0]$), and the third minimum (state
    $[1+1_3|0]$). The $\gamma(u)$ curve for the $[1|0]$
    state is shown for comparison. The phase-gradient profiles  
    corresponding to the bias points A through D are shown
    in Fig.~\ref{Fig:Profiles[2|0]}.}
  \label{Fig:IVC[2|0]}
\end{figure}
\begin{figure}
  \caption{
    The profiles of the phase gradients $\phi^{A,B}_x(x)$ in
    the $[2|0]$ states at the bias points A through D marked in
    Fig.~\ref{Fig:IVC[2|0]}.
  }
  \label{Fig:Profiles[2|0]}
\end{figure}
\begin{figure}
  \caption{
    (a) Current-velocity characteristics of the bunched
    states $[4|0]$, $[3|0]$, and $[2+1|0]$. Phase profiles
    of $[4|0]$ state and $[3|0]$ state at $\gamma=0.3$ are
    shown in (b) and (c), respectively.
  }
  \label{Fig:IVC[2+2|0]}
\end{figure}
%
%
\begin{figure}
  \caption{
    Current-velocity characteristics of the state $[0|1|0]$,
    bunched state $[0|1+1_M|0]$ for three different cases,
    $M=1,2,3$, and the state $[0|3|0]$. The profiles of
    the Josephson phase gradients at the points A through D are
    shown in Fig.~\ref{Fig:Profiles[0|2|0]}
  }
  \label{Fig:IVC[0|2|0]}
\end{figure}
\begin{figure}
  \caption{
    The profiles of the Josephson phase gradients
    $\phi^{A,B}_x(x)$ in $[0|1+1_M|0]$ states at the
    points A through D marked in Fig.~\ref{Fig:IVC[0|2|0]}.
  }
  \label{Fig:Profiles[0|2|0]}
\end{figure}


\begin{thebibliography}{99}

\bibitem[*]{gold-mail}
  e-mail: e.goldobin@fz-juelich.de , homepage: http:\BR{//}www\BR{.}geocities\BR{.}com\BR{/}e\_goldobin

\bibitem[\dag]{Boris-mail}
  e-mail: malomed@eng.tau.ac.il

\bibitem{Intrinsic}
  R.~Kleiner, F.~Steinmeyer, G.~Kunkel, and P.~M\"{u}ller,
  Phys. Rev. Lett. {\bf 68}, 2394 (1992);\\
  %
  R.~Kleiner and P.~M\"{u}ller,
  Phys. Rev.~B {\bf 49}, 1327 (1994).

\bibitem{Kl-Mue-Kolh:94}
  R.~Kleiner, P.~M\"{u}ller, H.~Kohlstedt,
  N.~F.~Pedersen, and S.~Sakai,
  Phys. Rev.~B {\bf 73}, 3942 (1994).

\bibitem{Paul:AdvSSP34}
  P.~M\"uller,
  In: {\it Festk{\"o}rperprobleme Advances in Solid State
  Physics},
  vol. {\bf 34}, ed. by Helbig (Vieweg, Braunschweig), p.~1
  (1995).

\bibitem{pl:pap154}
  A.~V.~Ustinov.
  In: {\it Physics and Materials Science of Vortex States,
  Flux Pinning and Dynamics}, NATO Science Series E, vol.
  {\bf 356}, edited by R.~Kossowsky \etal, Kluwer Acad.
  Publ. (1999), pp.~465--488.

\bibitem{image}
  Yu.~S.~Kivshar and B.~A.~Malomed, 
  Phys. Rev. B {\bf 37}, 9325 (1988).

\bibitem{Cherry1}
  E.~Goldobin, A.~Wallraff, N.~Thyssen, and A.~V.~Ustinov,
  Phys. Rev. B {\bf 57}, 130 (1998).

\bibitem{Cherry2}
  E.~Goldobin, A.~Wallraff, and A.~V.~Ustinov,
  accepted to J. Low Temp. Phys. (Nov 1999).
  see also cond-mat/9910234

\bibitem{HexFisherCherry}
  G.~Hechtfischer, R.~Kleiner, A.~V.~Ustinov and
  P.~M\"uller,
  Phys. Rev. Lett. {\bf 79}, 1365 (1997).

\bibitem{Ustinov:BunchArray}
  A.~V.~Ustinov, B.~A.~Malomed, and S.~Sakai,
  Phys. Rev. B {\bf 57}, 11691 (1998).

\bibitem{Sakai:BunchBeta}
  S.~Sakai,
  Phys. Rev. B {\bf 36}, 812 (1987).

\bibitem{Malomed:BunchBeta}
  B.~A.~Malomed,
  Phys. Rev. B {\bf 47}, 1111 (1993).

\bibitem{Vernik:BunchBeta}
  I.~V.~Vernik, N.~Lazarides, M.~P.~S\o{}rensen,
  A.~V.~Ustinov, N.~F.~Pedersen, and V.~A.~Oboznov,
  J.~Appl. Phys. {\bf 79}, 7854 (1996).

\bibitem{SBP}
  S.~Sakai, P.~Bodin, and N.~F.~Pedersen.
  J.~Appl. Phys. {\bf 73}, 2411 (1993).

\bibitem{LT21}
  E.~Goldobin, A.~Golubov, A.~V.~Ustinov,
  Czech. J. Phys. {\bf 46}, 663 (1996), LT-21 Suppl. S2

\bibitem{StkJJ}
  E.~Goldobin, available online:
  http:\BR{//}www\BR{.}geocities\BR{.}com\BR{/}SiliconValley\BR{/}Heights\BR{/}7318\BR{/}StkJJ.htm
  (1999).

\bibitem{GrE:Stability}
  N.~Gr{\o}nbech-Jensen, D.~Cai and M.~R.~Samuelsen,
  Phys. Rev. B {\bf 48}, 16160 (1993).

\end{thebibliography}
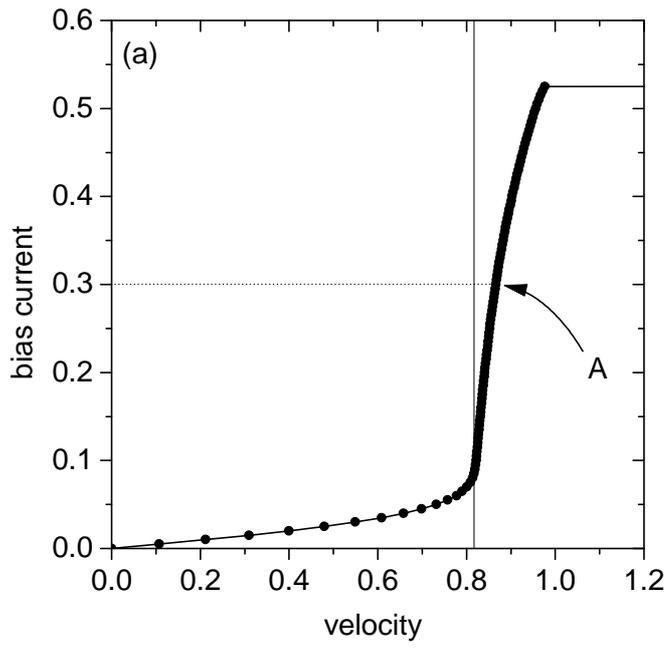
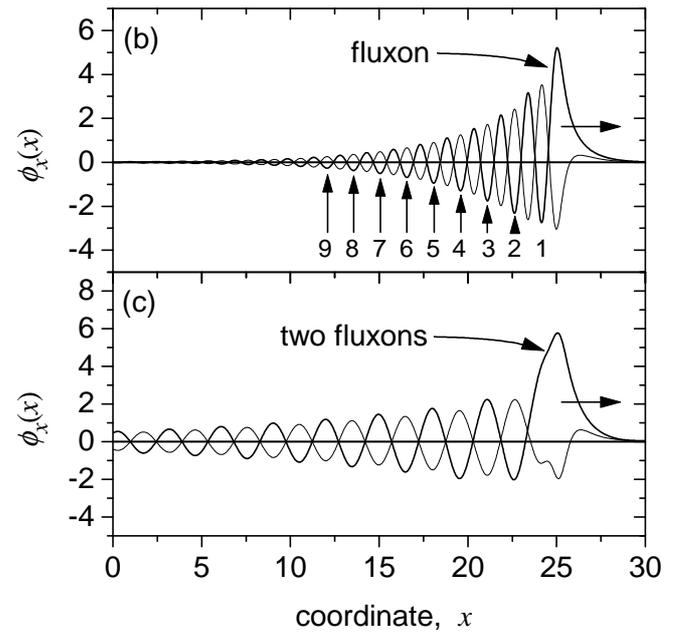
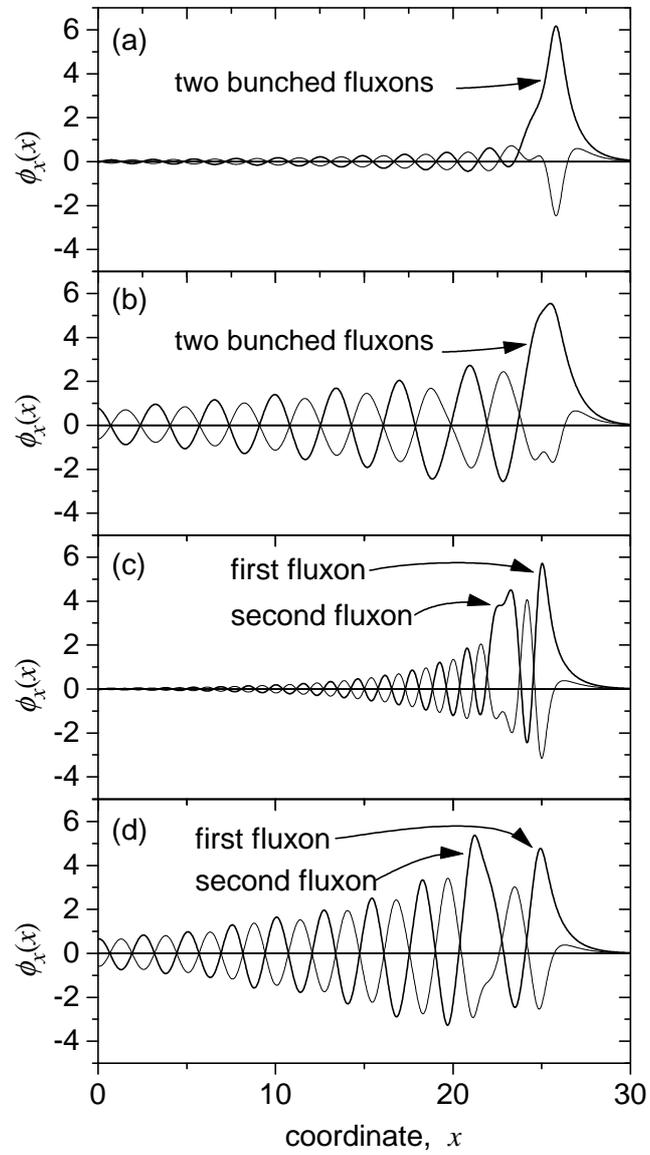
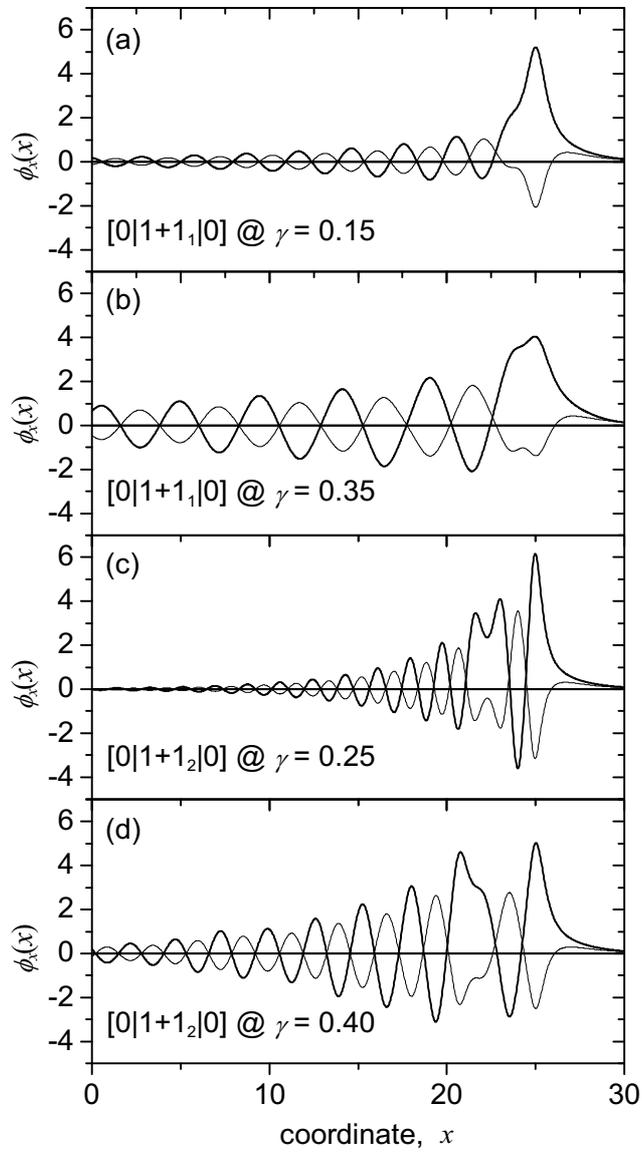
\end{document}